\documentclass[journal]{IEEEtran}  %onecolumn
\usepackage[pdftex]{graphicx}
\usepackage{amssymb}
\usepackage{amsmath}
\usepackage{algorithmic}
\usepackage{array}
\usepackage[style=ieee, doi=false, isbn=false, url=false, eprint=false, maxnames=3]{biblatex}
\usepackage{xcolor}
\usepackage[caption=false,font=footnotesize]{subfig}
\usepackage{framed}  % put frames around figures \begin{framed} \end{framed}
\usepackage{setspace} 
\usepackage[colorlinks=true, linkcolor=blue, anchorcolor=blue, citecolor=blue]{hyperref}

\begin{document}
\title{\huge Designing a Self-Decoupled 16 Channel Transmitter for Human Brain Magnetic Resonance Imaging at 447MHz}

\author{Nader Tavaf, Jerahmie Radder, Russell L. Lagore, Steve Jungst, Andrea Grant, \\Kamil Ugurbil, Gregor Adriany, Pierre Francois Van de Moortele
\thanks{Center for Magnetic Resonance Research, University of Minnesota Twin Cities. Minneapolis, MN 55455. E-mail: tavaf001@umn.edu.}
%\thanks{}
}

\markboth{Tavaf \MakeLowercase{\textit{et al.}}: Self Decoupled 16 Channel Transmitter}%
{}

\maketitle

\begin{abstract}
Transmitter arrays play a critical role in ultra high field Magnetic Resonance Imaging (MRI), especially given the advantages made possible via parallel transmission (pTx) techniques. One of the challenges in design and construction of transmit arrays has traditionally been finding effective strategies for decoupling elements of the transmit array. Here, we present the design of the first self-decoupled, loop-based transmit array for human brain MRI at 10.5T /447MHz. We demonstrate, using full-wave electromagnetic simulations, effective decoupling of the transmit elements without requiring the conventional overlap or inductive decoupling techniques.
\end{abstract}

\section{Introduction}
\IEEEPARstart{M}{agnetic} Resonance Imaging (MRI) relies on radio frequency (RF) antenna arrays for transmission and reception of the signal. At ultra high field (UHF), defined as those MRI systems with static magnetic field $B_0\ge 7$Tesla (T), the higher frequency / shorter wavelength provides significant opportunities~\cite{Ugurbil2014, Ugurbil2018} while it complicates the design and construction of radio-frequency arrays~\cite{Adriany2019, Tavaf2020a}. The interaction between the elements of the transmit array can distort the individual magnetic field $B_1$ maps. Furthermore, the current induced on the coaxial cables feeding the transmit elements can also interfere with the field maps. These complications pose challenges to the safety validation of transmit arrays, where the combined magnetic field of the array measured using phantom experiments must be replicated in full-wave electromagnetic / circuit co-simulations with such accuracy as is necessary for ensuring the safety of prospective human subjects. The match between the measured and simulated magnetic field is required to guarantee that the accuracy of the specific absorption rate (SAR) calculations which rely on the simulated electric fields. 

In our previous work \cite{Adriany2019}, we presented a 16-channel dual row, loop transmit array for human brain imaging at 10.5T. Previous transmit arrays relied on overlap and/or inductive decoupling between elements~\cite{Adriany2019, Avdievich2018a, Shajan2016, Shajan2014}. Furthermore, the interaction between the transmit elements is in practice further complicated by the currents induced on the coaxial cables. Such challenges motivated us to try to simplify the design of the transmit array, primarily by removing the inductive (transformer) decoupling (as hand-wound transformers are difficult to replicate accurately in simulation). To do so, we had to rely on previous studies demonstrating effective self-decoupling of radio frequency coil elements~\cite{Lakshmanan2014, Yan2018, Tavaf2020a}. Self-decoupling in essence relies on unbalanced current distribution in each loop element to mitigate interaction between adjacent coils without requiring overlap or other decoupling strategies. Here, we present the first self-decoupled 16-channel transmit array and demonstrate, via full-wave electromagnetic/circuit co-simulations, that (1) the self-decoupling provides sufficient isolation allowing for the construction of the transmit array, and importantly, (2) the individual field maps of each transmit element are free of distortions. These initial results indicate a promising path to building a transmitter that would facilitate the safety validation process. Furthermore, we share with the public the simulation model, the data and the analysis code. These can serve the research community well by providing an annotated guideline to the simulation and analysis of radio frequency arrays for MR applications as well as an opportunity to replicate these results.

\section{Methods}
We modeled 16 channels, each being a 10x10 cm\textsuperscript{2} square loop with a thickness of 0.05 mm (representative of copper tape). These square loops were segmented at their four corners (i.e. vertices). The 16 loops are wrapped around in a cylindrical surface of diameter 285mm, and are distributed in two rows of eight loops each, with the second row having a 22.5 degree azimuthal rotation (around the \textit{z}-axis) compared to the first row. There is a gap of approximately 10mm between the loops, both between the loops from different rows (along the \textit{z}-axis) and between the loops within the same row (along the azimuthal perimeter). The capacitor at the feed point was a trimmer in the range of 5-10 pF, while opposite the feed, another trimmer tuned to 0.5 pF was used. The other two segmenting capacitors were approximately 1.7 pF. The loops are not overlapped and there is no inductive decoupling involved in this transmit array. \autoref{fig:head_model} shows the arrangement of the loops around a human head-shaped phantom.

\subsection{Phantoms}
\begin{figure}
    \centering
    \subfloat[]{\includegraphics[width=0.5\linewidth]{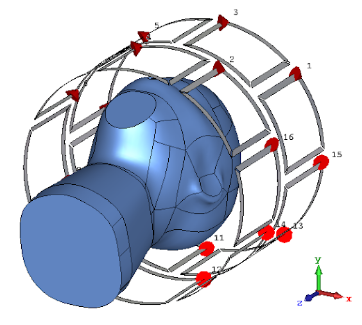}}
    \subfloat[]{\includegraphics[width=0.5\linewidth]{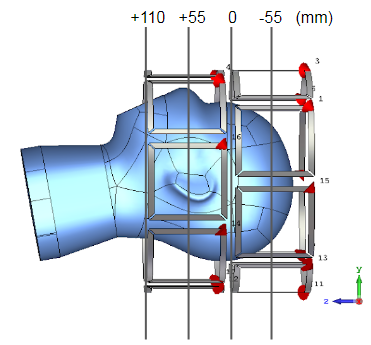}}
    \caption{The simulation model, showing the human head shaped phantom. (a) the perspective view. (b) the view from the patient's left side, showing the location of the axial cross sections used in the field maps.}
    \label{fig:head_model}
\end{figure}

Two phantoms were used in our simulations, a cylindrical phantom of diameter 245mm, and a human head shaped phantom. The dielectric material properties of these phantoms were set to $\epsilon=49.7, \sigma=0.66$ S/m to reflect the properties of the human brain tissue~\cite{Hasgall2018} and the dielectric properties of the phantoms we had built in our research center (which will later be used for experimental validation). \autoref{fig:head_model} shows the model using the human head shaped phantom.

The reason for using two different phantoms was to isolate the complications arising from the asymmetric shape of the human head. In other words, analyzing the cylindrical phantom removes such complication and makes it straightforward to see the effects of self-decoupling in the absence of other factors complicating the analysis.

\subsection{Simulations}
Electromagnetic simulations were performed in CST Studio (Dassault Systèmes Simulia Corp., Johnston, RI). Simulations were performed over a frequency range of 1 GHz using the finite-difference time-domain method to solve Maxwell’s equations. Details of the simulation pipeline are discussed in our previous work~\cite{Adriany2019}.

\subsection{Data Analysis}
The current density and S-parameters were directly exported from CST. The magnetic fields exported from CST were used to calculate the transmit field maps using \[B_1^+ = \mu_0 (H_x + 1j \cdot H_y)/2\] where $\mu_0$ is the magnetic permeability in vacuum, and $H$ is the magnetic field exported from CST. These calculations were performed for individual channels for individual slices, albeit in matrix form. The circularly polarized (``CP") mode for combining the field maps is calculated by adding incrementally increasing phase to the field maps of these 16 channels (in $2\pi /16$ increments). The ``CP2+" mode for combining transmit field maps is calculated similarly but using $4\pi/16$ phase increments. The summation over channels is carried out in complex domain, but the results presented in the next section show only the magnitude of the field maps. These particular combinations of the complex field maps are of interest because we know what to expect from their results and can compare them with existing similarly calculated results from previous studies. 

\begin{figure}[t]
    \centering
    \subfloat[]{\includegraphics[width=0.45\linewidth]{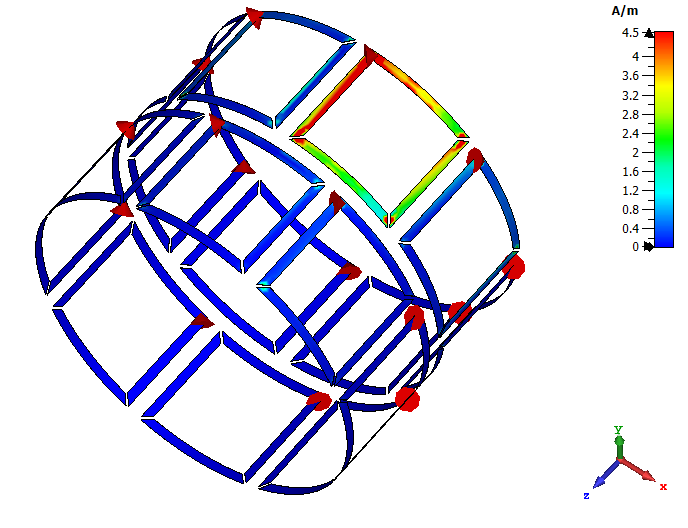}}
    \hfill
    \subfloat[]{\includegraphics[width=0.45\linewidth]{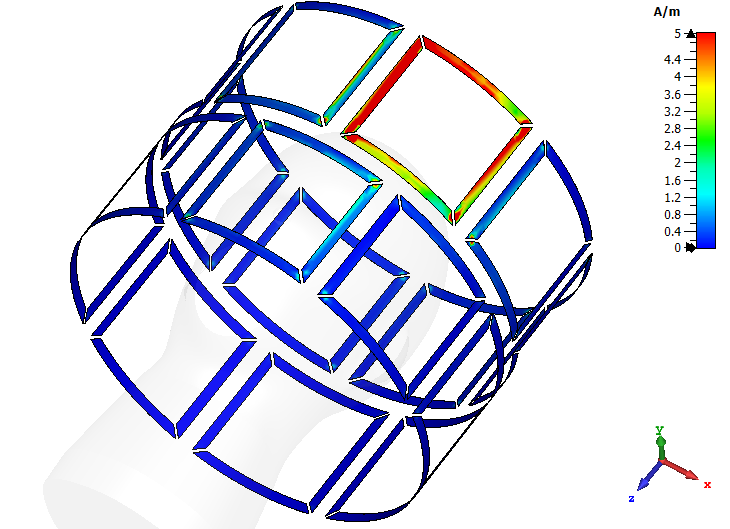}}
    \caption{Current density of a single loop when only that loop is excited (a) using the cylindrical phantom, (b) with the human head shaped phantom.}
    \label{fig:current}
\end{figure}
\begin{figure}[t]
    \centering
    \subfloat[]{\includegraphics[width=0.4\linewidth]{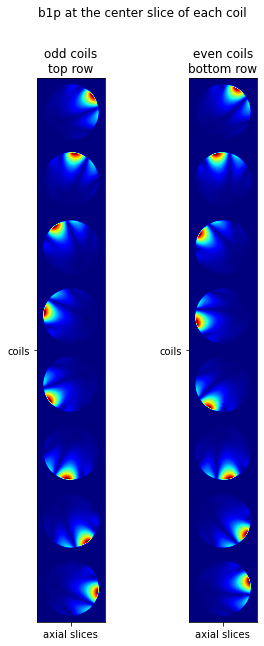}}
    \hspace{0.05\linewidth}
    \subfloat[]{\includegraphics[width=0.4\linewidth]{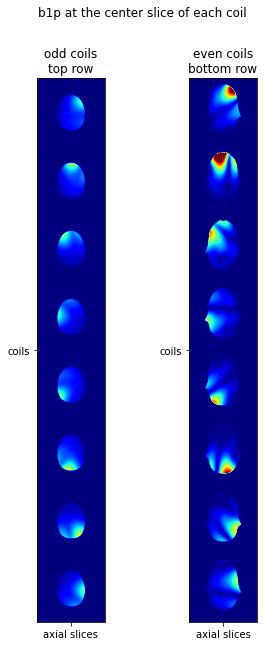}}
    \caption{Individual field maps of each channel at the axial slice crossing the center of that channel (a) using the cylindrical phantom, (b) with the human head shaped phantom.}
    \label{fig:individual_field}
\end{figure}

\section{Results}
\begin{figure*}
    \centering
    \subfloat[]{\includegraphics[height=3cm]{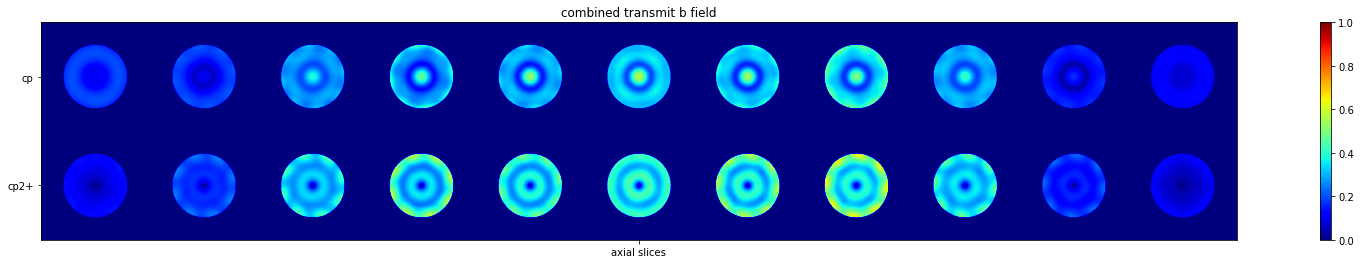}}
    \vfill
    \subfloat[]{\includegraphics[height=3cm]{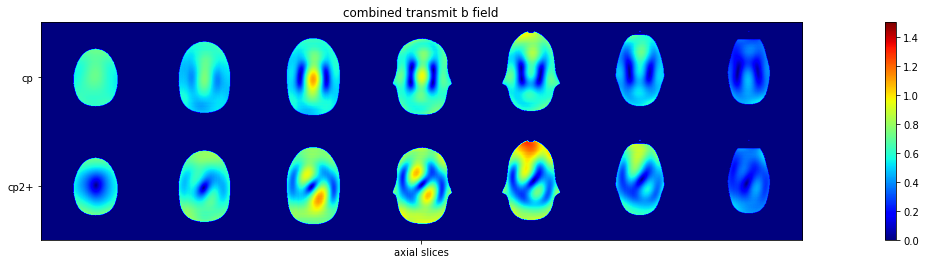}}
    \caption{Combined field maps of the 16 channels at several axial slices (a) using the cylindrical phantom, (b) with the human head shaped phantom. The top row is the CP mode combination, the bottom row is the CP2+ mode combination. In (b) with the human head shaped phantom, the slices are at \textit{z}=[-55, -27.5, 0, +27.5, +55, +82.5, +110] (from left to right), such that the third slice from left is the central slice between the two rows of loops.}
    \label{fig:combined_field}
\end{figure*}
S-parameters of the coil using both phantoms are displayed in the Appendix. With the cylindrical phantom, the $S_{ii}$ parameters from the simulations are on average -15dB, the coupling between the neighboring elements from different rows is better than -18dB in all cases, and the $S_{ij}$ between the neighboring elements inside the same row is better than -14dB in all cases. These results indicate that self-decoupling design provides sufficient isolation between channels. Notice that no preamplifier decoupling is involved in these simulations. The human head shaped phantom, however, is asymmetric and does not load the top row of the coil as much, resulting in some residual coupling between the elements of the top row even though the $S_{ii}$ do not show split resonances.

The current density maps provided in \autoref{fig:current} indicate that while there is strong current on one of the edges parallel to the \textit{z} axis of the loop, there is not much current on the other edge parallel to the \textit{z} axis. More importantly, there is little current induced on the edge shared with the next row of loops (i.e. the edge perpendicular to the \textit{z} axis). These unbalanced current distributions help with mitigating crosstalk, as is evident from the limited current induced on the adjacent coils when one loop element is excited. This holds true in both the cylindrical and human head shaped case.

The individual field maps are presented in \autoref{fig:individual_field}. In either \autoref{fig:individual_field}(a) or \autoref{fig:individual_field}(b), the left column is the field map from the eight channels at the top row whereas the right column is the field maps from each of the eight channels at the bottom row. In the case of the cylindrical phantom, the field maps are characteristic of loop elements and are not distorted. With the asymmetric human head shaped phantom, the phantom is further away and does not heavily load the top row of the coils, as is evident from the low intensity of the $B_1$ strength for these coils. Nevertheless, even with the asymmetric load, the field maps still maintain their characteristic two-lobe shape, with the stronger lobe corresponding to the vertical edge of the loop with stronger current. The absence of distortions or extraneous lobes in the field intensity profiles corroborates the decoupling of the coils.

The combined field maps are presented in \autoref{fig:combined_field}. \autoref{fig:combined_field}(a) which reflects the cylindrical case, is particularly well-behaved, with its CP mode consistent with our previous experimental results~\cite{Adriany2019}, and the CP2+ mode as expected. \autoref{fig:combined_field}(b) presents combined field maps in axial slices with the human head shaped phantom. Note that we are not shimming the field maps for either homogeneity or efficiency, this is rather explicitly the CP and CP2+ combination of the field maps. The CP mode results with the human head shaped phantom are also consistent with previous results (see for example figure 3 of \cite{Avdievich2018a}).

\section{Conclusion and Discussion}
The effective decoupling of the elements of a transmit array is critical to ensuring the absence of artificial distortions of the field maps of individual channels especially in ultra high field.
Here, we presented the design of the first self-decoupled 16 channel transmitter array for human brain imaging at 10.5T. Compared to previous works, the exclusion of inductive decoupling simplifies the design and construction of the array. This is a significant milestone as it is expected to facilitate the safety validation of the transmit arrays. 
%Beyond the use of self-decoupling in design of the transmitters, fundamental paradigm shifts in the safety validation process, such as deep learning-based techniques~\cite{Meliado2019, Tavaf2021_grappa_gan}, can further streamline the process, for instance by complementing the full-wave electromagnetic simulations.
In continuing this work, we are constructing the transmit array and anticipate validation of these results via phantom experiments at 10.5T.

\AtNextBibliography{\small}
\printbibliography

\appendices
\section{Scatter Matrix Parameters}
The S parameter figures are presented here in the appendix.
\begin{figure*}
    \centering
    \subfloat[]{\includegraphics[width=\textwidth]{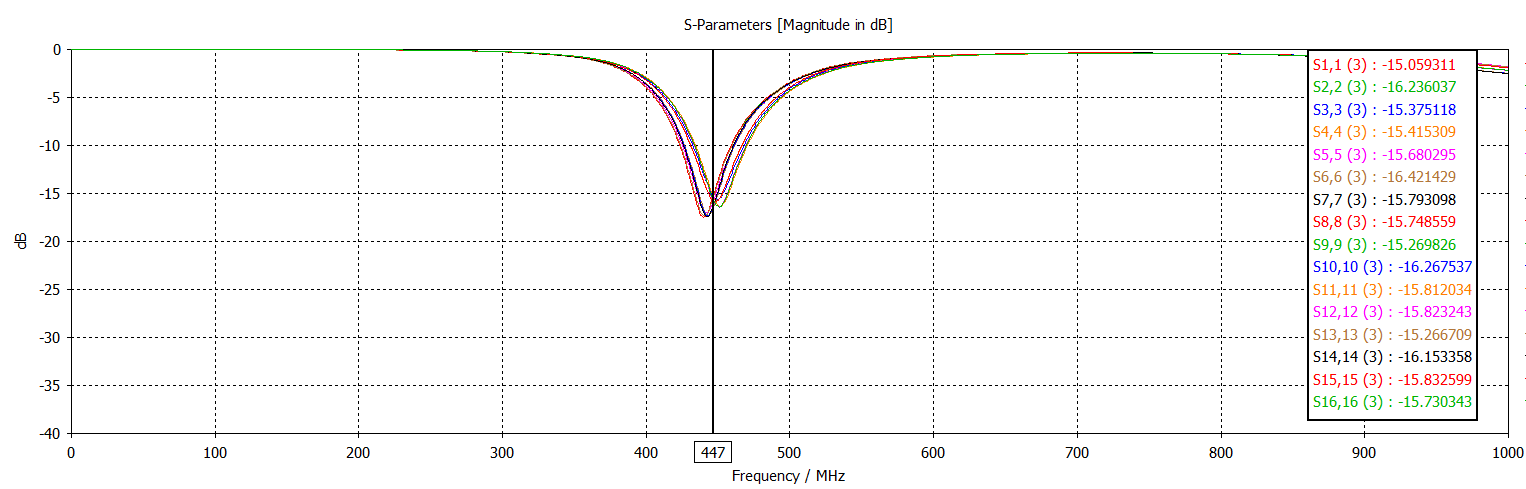}}
    \vfill
    \subfloat[]{\includegraphics[width=0.45\textwidth]{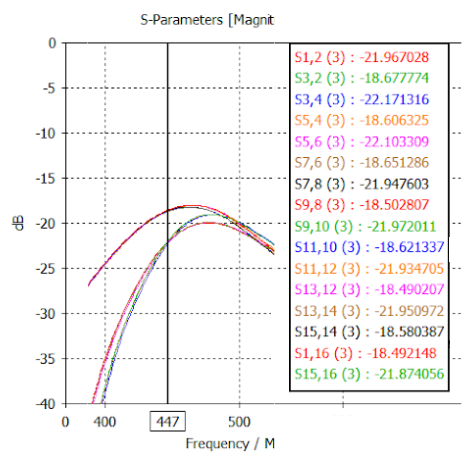}}
    \hfill
    \subfloat[]{\includegraphics[width=0.44\textwidth]{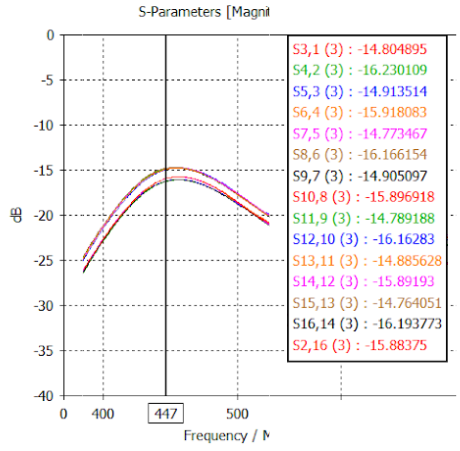}}
    \caption{The S parameters for all of the channels with the cylindrical phantom. (a) $S_{ii}$ for each channel. (b) $S_{ij}$ between neighboring elements of different rows. (c) $S_{ij}$ between neighboring elements within the same rows. }
    \label{fig:cylinder_s_params}
\end{figure*}
\begin{figure*}
    \centering
    \subfloat[]{\includegraphics[width=\textwidth]{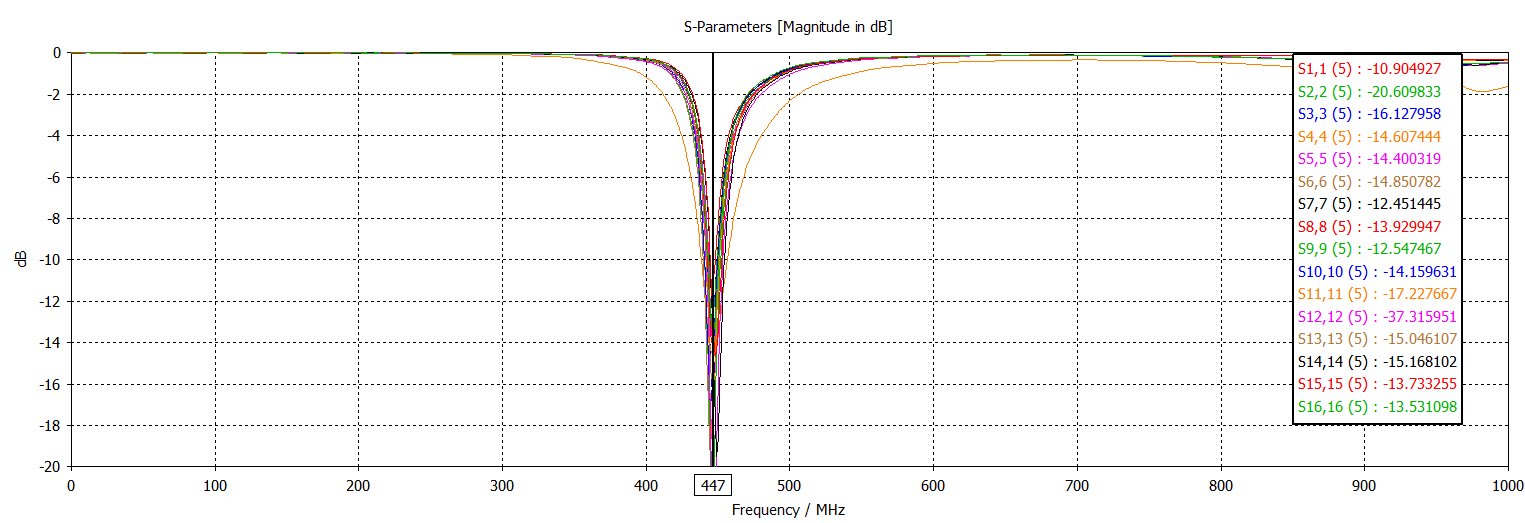}}
    \vfill
    \subfloat[]{\includegraphics[width=0.43\textwidth]{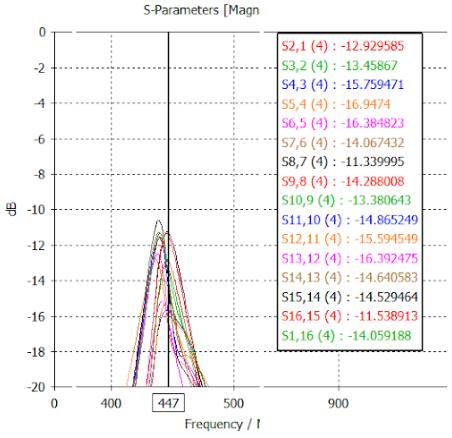}}
    \hfill
    \subfloat[]{\includegraphics[width=0.45\textwidth]{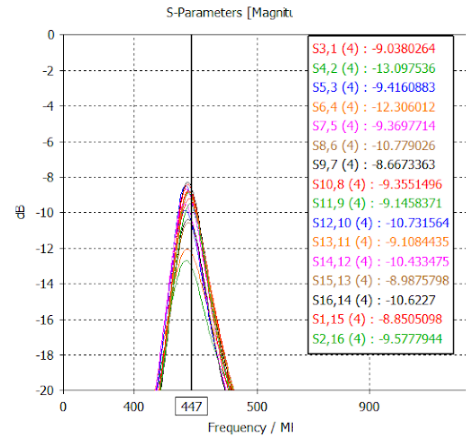}}
    \caption{The S parameters for all of the channels with the head shaped phantom. (a) $S_{ii}$ for each channel. (b) $S_{ij}$ between neighboring elements of different rows. (c) $S_{ij}$ between neighboring elements within the same rows. }
    \label{fig:head_s_params}
\end{figure*}
\end{document}